\documentclass[journal=jcisd8,manuscript=article]{achemso}
\usepackage[range-units=single]{siunitx}
\DeclareSIUnit{\calorie}{cal}
\usepackage{graphicx}
\usepackage{bm}
\usepackage{amsmath}
\usepackage{amssymb}
\usepackage{hyperref}
\usepackage{booktabs}
\usepackage{mathtools}
\usepackage{csquotes}
\usepackage{diffcoeff}
\usepackage[version=4]{mhchem}
\usepackage[dvipsnames]{xcolor}
\DeclareMathOperator{\VJP}{VJP}
\DeclareMathOperator{\SoftPlus}{SoftPlus}
\DeclareMathOperator{\erf}{erf}
\DeclareMathOperator{\VACF}{VACF}
\DeclarePairedDelimiter{\pqty}{(}{)}
\DeclarePairedDelimiter{\sqty}{[}{]}
\DeclarePairedDelimiter{\bqty}{\lbrace}{\rbrace}
\DeclarePairedDelimiter{\aqty}{\langle}{\rangle}

\DeclareSIUnit\atm{atm}
\DeclareSIUnit\atom{atom}
\DeclareSIUnit\calorie{cal}
\newcommand{\onlinecite}[1]{\hspace{-1 ex} \nocite{#1}\citenum{#1}} 

\usepackage[T1]{fontenc}

\title{A Differentiable Neural-Network Force Field for Ionic Liquids}

\author{Hadri\'an Montes-Campos}
\altaffiliation{Contributed equally to this work}
\affiliation[University of Santiago de Compostela]{Grupo de Nanomateriais, Fot\'onica e Materia Branda, Departamento de F\'isica de Part\'{\i}culas, Universidade de Santiago de Compostela, Campus Vida s/n E-15782, Santiago de Compostela, Spain}

\author{Jes\'us Carrete}
\email{jesus.carrete.montana@tuwien.ac.at}
\altaffiliation{Contributed equally to this work}
\affiliation[TU Wien]{Institute of Materials Chemistry, TU Wien, 1060 Vienna, Austria}

\author{Sebastian Bichelmaier}
\affiliation[TU Wien]{Institute of Materials Chemistry, TU Wien, 1060 Vienna, Austria}

\author{Luis M. Varela}
\affiliation[University of Santiago de Compostela]{Grupo de Nanomateriais, Fot\'onica e Materia Branda, Departamento de F\'isica de Part\'{\i}culas, Universidade de Santiago de Compostela, Campus Vida s/n E-15782, Santiago de Compostela, Spain}

\author{Georg K. H. Madsen}
\affiliation[TU Wien]{Institute of Materials Chemistry, TU Wien, 1060 Vienna, Austria}

\begin{document}

\begin{tocentry}
    \includegraphics{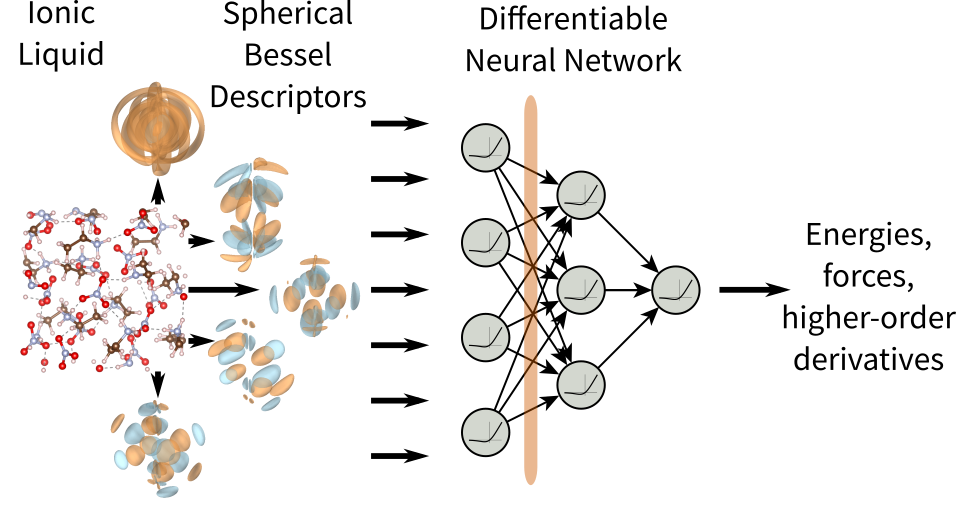}
\end{tocentry}

\begin{abstract}
    We present \textsc{NeuralIL}, a model for the potential energy of an ionic liquid that accurately reproduces first-principles results with orders-of-magnitude savings in computational cost. Based on a multilayer perceptron and spherical Bessel descriptors of the atomic environments, \textsc{NeuralIL} is implemented in such a way as to be fully automatically differentiable. It can thus be trained on ab-initio forces instead of just energies, to make the most out of the available data, and can efficiently predict arbitrary derivatives of the potential energy. Using ethylammonium nitrate as the test system, we obtain out-of-sample accuracies better than \SI{2}{\milli\electronvolt\per\atom} ($<\SI{0.05}{\kilo\calorie\per\mol}$) in the energies and \SI{70}{\milli\electronvolt\per\angstrom} in the forces. We show that encoding the element specific density in the spherical Bessel descriptors is key to achieving this. Harnessing the information provided by the forces drastically reduces the amount of atomic configurations required to train a neural network force field based on atom-centered descriptors. We choose the Swish-1 activation function and discuss the role of this choice in keeping the neural network differentiable. Furthermore, the possibility of training on small data sets allows for an ensemble-learning approach to the detection of extrapolation. Finally, we find that a separate treatment of long-range interactions is not required to achieve a high-quality representation of the potential energy surface of these dense ionic systems.
\end{abstract}

\section{Introduction}
\label{sec:introduction}

Room-temperature ionic liquids\cite{macfarlane2017fundamentals} (ILs) are ionized substances that exist in the liquid state at temperatures below \SI{100}{\celsius}. Two broad classes can be defined: protic ILs, which are formed by proton transfer from an acid to a base, and aprotic ILs based on an organic molecular cation and an anion that can range from a single atom to another complex structure. ILs are very interesting from a fundamental point of view because of the many peculiar features of their dynamics, arising from the competition of electrostatic, steric and dispersion interactions among ions, but their prominence in the scientific literature is undoubtedly mostly due to their potential for applications in industry \cite{surface_electrochemistry}. ILs as a class have some desirable properties in this regard, the best known one being the negligible vapor pressure of aprotic ILs that makes it possible to use them as \enquote{green solvents} \cite{green_solvents} free from leaks to the environment. However, their greatest promise lies in their diversity: a million binary and a quintillion ternary ILs are theoretically possible through the choice of anions and cations, compared to the $\sim600$ organic solvents in current use \cite{Rogers792}. Amid that vast landscape, compounds have been found that fulfill specific requirements such as stability (e.g. against thermal decomposition\cite{review_thermal_stability} or in mixtures with water\cite{review_water_mixtures}), biocompatiblity\cite{review_biocompatible} or wide electrochemical windows\cite{review_electrochemical_window}. It is therefore plausible that tailored ILs could be found for many applications, leading to the inclusion of ILs under the label of \enquote{designer solvents} as well.

Unfortunately, that aprioristic enthusiasm has to coexist with the fact that brute-force exploration of the possible ILs is inconceivable. In this context, computer modeling and simulation are invaluable complements to experiment, providing insight into the connection between structure and functionality at the atomic level and suggesting new substances to explore. However, to properly reproduce the structural and dynamical correlations in a liquid, a significant quantity of substance has to be included in a simulation, and the atomic trajectories have to be traced for times of the order of nanoseconds or longer. Consequently, ab-initio molecular dynamics studies are usually limited to those phenomena that can be understood in terms of the fine details of the behavior of a few ionic pairs \cite{aimd2007,aimd2018}.

Classical molecular dynamics (MD) simulations are a better fit for the scales of time and quantity of substance required and have been used extensively to study pure ILs and their mixtures \cite{il_md1, il_md2, il_md3, il_md4}. However, abandoning ab-initio methods incurs a high cost in terms of accuracy and transferability. The centerpiece of an MD simulation of an IL is a molecular-mechanics force field (FF), of which OPLS-AA is a very representative example \cite{opls_aa, opls_aa_il}. While OPLS-AA contains a large number of parameters, they are easily interpretable and can be systematically fitted to modest amounts of ab-initio data, eliminating the need for a prohibitively costly global fit. However, the predictions of molecular-mechanics FFs have qualitative rather than quantitative value. An improvement over plain molecular-mechanics FFs comes from polarizable FFs \cite{polarizable_review}, which try to add some flexibility by allowing an induced dipole moment to appear at each atom in reaction to the local electric field. The effect of polarizability has been compared to that of a solvent \cite{polarizable_review}, making the predicted structure and dynamics less similar to those of an ionic solid. As an example, the predicted structural properties of 1-ethyl-3-methylimidazolium bis-(trifluoromethylsulfonyl)-imide doped with a lithium salt barely change when switching to a polarizable FF, but the diffusion coefficients can change by up to an order of magnitude \cite{polarizable_lithium}. The pinnacle of molecular mechanics can be considered to be ReaxFF \cite{reaxff_review}, a reactive FF with a variable topology, a difficult parametrization process and terms inspired by quantum chemistry. Still, even ReaxFF has run up against the limitations of \enquote{physically inspired} building blocks and has been forced to branch into specialized parametrizations.

Recently, a completely different approach to the understanding and development of FFs has emerged in the context of machine learning (ML) in computational chemistry. The parametrization of an FF is regarded as a regression problem, where a set of continuous inputs (Cartesian coordinates) must be mapped to a set of continuous outputs (energies and forces) in an optimal manner. The focus is hence shifted towards a sufficiently general functional form that can be efficiently trained on the available data. While alternatives exist, such as Gaussian process regression \cite{bartok_2015} and the more recent Euclidean neural networks (NNs) \cite{Batzner_21}, one of the most fertile approaches to constructing MLFFs is based on fully-connected NNs following a general template where the total energy is constructed as a sum of atomic energies \cite{behler_tutorial,behler_review_2021, optimizing_nn_potentials}. To preserve the fundamental symmetries of mechanics, the atomic energies depend on the local chemical environment through explicitly scalar atom-centered descriptors, rather than directly on the Cartesian coordinates.

In this paper we present \textsc{NeuralIL}, an NNFF for ILs based on atom-centered descriptors. We train and apply it to the IL ethylammonium nitrate (EAN) and show that the results offer quality comparable to first-principles calculations at a small fraction of the cost. The NNFF uses the second-generation spherical Bessel descriptors introduced by Kocer \textit{et al.} \cite{kocer2020continuous} Compared to the more widely used atom-centered symmetry functions \cite{Behler_JCP11} and the smooth overlap of atomic positions \cite{Bartok_PRB13} descriptors, the spherical Bessel descriptors have been shown to minimize the amount of redundant information in the expansion\cite{kocer2020continuous}. We generalize the spherical Bessel descriptors so that they do not rely on arbitrary weights for the different elements and show that this generalization is essential to precisely model the ab-initio data. Furthermore, to fully capture the chemical nature of the atoms, the descriptors are augmented with an embedding vector.

\textsc{NeuralIL} puts special emphasis on the forces. Through careful implementation choices, we show how the full data pipeline, from the Cartesian coordinates to the model, can be made automatically differentiable \cite{ad_in_ml}. Thereby our model can predict forces efficiently and can also be trained on them, making optimal use of the data obtained from the ab-initio calculations. Compared to using only the total energy, where just one data point is obtained per atomic structure, $3n_\textrm{atoms}$ force components are routinely provided by ab-initio calculations.

Automatic differentiation is a key piece of the modern ML landscape \cite{ad_in_ml}. As far as interatomic potentials are concerned, automatic differentiation often plays a role in equivariant convolutional NN models \cite{SchNet, Cormorant, Batzner_21} but has yet to be widely introduced for descriptor-based NNFFs \cite{TorchANI}. Automatic differentiation makes workarounds such as local Taylor approximations \cite{taylor_expansion} or atomic decomposition \cite{energy_decompositions} of DFT energies unnecessary. \enquote{Hands-off} training of NNFFs is typically based on datasets ranging from hundreds of thousands to millions of atomic configurations \cite{natalio_anharmonic_nn,ani1}. The present design and the possibility to train on forces do away with the idea that these large databases are required for descriptor-based NNFFs. At the same time, descriptor-based NNFFs still guarantee that a potential energy consistent with the forces exists (i.e., that forces are conservative), which cannot be taken for granted if the forces are regarded as an arbitrary vector field during training.

The next section contains the details of the descriptors, the NN, the implementation and the training procedure. Then we analyze the results for EAN, discuss their implications for the model in general, and provide some comparisons with other ways to encode the chemical information. We furthermore show that a sufficiently flexible and accurate short-range potential provides a perfectly satisfactory description of this IL, and that a molecular-mechanics-inspired treatment of Coulomb interactions in terms of static atomic charges in fact degrades the results. We also devise and demonstrate an inexpensive method to assess the transferability to the trained model to a new point in configuration space by using an ensemble of NNs. Finally, we summarize our main conclusions.

\section{Methods}
\label{sec:methods}

\subsection{Ab-initio calculations}
\label{subsec:training}
The database of EAN configurations created for this work is provided as part of the supplementary material. The main use case of FFs for ILs is to obtain improved results for MD simulations under conditions close to room temperature. Therefore, our data set is built on the basis of configurations sampled from a classical MD trajectory, which are then treated using density functional theory (DFT).

To run the classical MD simulations, we use \textsc{Gromacs} \cite{GROMACS} with the OPLS-AA FF\cite{opls_aa, opls_aa_il}. The details of our parametrization of EAN are given in Ref.~\onlinecite{opls_parametrization}. Our starting point is a cubic box with a side length of \SI{1.29}{\nano\meter}, filled with $15$ ionic pairs in order to achieve a density similar to that of the pure IL. The initial positions are generated with \textsc{Packmol} \cite{packmol} to avoid placing any pair of particles too close together. We then perform a conjugate-gradients minimization of the original coordinates, followed by a \SI{10}{\nano\second} stabilization run to bring the system to a reference temperature of \SI{298.15}{\kelvin} using a velocity-rescaling thermostat with a time constant of \SI{0.1}{\nano\second}. Finally, we run a \enquote{production} simulation of \SI{5}{\nano\second} starting from the stabilized box and store the resulting trajectory. All integrations are carried out using a velocity Verlet algorithm with a time step of \SI{1}{\femto\second}. A cutoff radius of \SI{0.6}{\nano\meter} is adopted for the long-range interactions. The van der Waals term is truncated at that distance, while Coulomb interactions are evaluated using a fast smooth particle-mesh Ewald method \cite{SPME}, with that same radius acting as an upper bound for the real-space term. Dispersion corrections are applied to the energy to account for the truncated van der Waals interaction in a mean-field approximation.

The \SI{5}{\nano\second} trajectory is subsampled to extract $741$ configurations. Each of those is used as an input to the \textsc{Gpaw} DFT package \cite{gpaw1,gpaw2} in linear-combination-of-atomic-orbitals (LCAO) mode \cite{Larsen_PRB09}, with a double-$\zeta$ plus polarization basis set, the local density approximation (LDA) to exchange and correlation (XC), a grid spacing of \SI{0.2}{\angstrom} and $\Gamma$-only sampling of the Brillouin zone. Since the DFT calculations are intended to generate a ground truth for the model, alternative XC parameterizations and semiempirical treatments of dispersive interactions are not explored. To improve sampling in areas close to the local minima of the ab-initio potential energy landscape, for $373$ of those configurations we then run the quasi-Newton minimizer implemented in \textsc{ASE} \cite{ase-paper} for five steps using the same DFT parameters. Using a fixed number of minimization steps helps avoid a situation where all initial configurations collapse around stationary points. The $373$ final structures after each minimization along with the $368$ remaining unminimized samples, each with their DFT energies and forces, make up the training data set.

\subsection{Atom-centered descriptors}
\label{subsec:descriptors}
The first step in constructing the NNFF is to transform the $3n_{\mathrm{atoms}}$ Cartesian coordinates of the system into a set of atom-centered descriptors. Those describe the atomic environments without encoding an absolute origin of coordinates or an absolute orientation of the axes and are thus explicitly translation- and rotation-invariant. Specifically, the quantity to be encoded is the local density of each chemical element $J$ around each atom $i$ in the system within a sphere of a predefined cutoff radius, which in the present study is set to a $r_c = \SI{3.5}{\angstrom}$,
\begin{equation}
    \rho_{iJ}\pqty*{\mathbf{r}} = \sum\limits_{\substack{j\in J\\R_{ij} < r_c\\j\ne i}} \delta\pqty*{\mathbf{r} - \mathbf{R}_{ij}}.
    \label{eqn:density}
\end{equation}
The cutoff radius was chosen after convergence tests for values up to $\SI{6.0}{\angstrom}$, which showed small improvements in accuracy with a significant impact on performance. Our descriptors are directly based on the density defined in Eq.~\eqref{eqn:density} for each chemical element $J$. In contrast, most earlier work (with some exceptions like Ref.~\onlinecite{ani1}) employ descriptors that encode one or more weighted densities of the form
\begin{equation}
    \rho_i\pqty*{\mathbf{r}} = \sum\limits_J w_J \rho_{iJ}\pqty*{\mathbf{r}}
    \label{eq:rhoweight}
\end{equation}
with predefined weights like atomic numbers or atomic masses. As explained below, under our approach the number of descriptors per atom increases quadratically with the number of chemical species in the system, while premixed densities make those two numbers independent. The  extensive comparisons between both possibilities reported in this article show that premixing leads to a significant loss of information in the descriptors and degrades the accuracy of the model.

Following the recipe for the spherical Bessel descriptors proposed by Kocer~\textit{et~al.}\cite{kocer2020continuous}, each density is projected on an orthonormal set of basis functions
\begin{equation}
    B_{n\ell m}\pqty{\mathbf{r}} = g_{n-\ell,\ell}\pqty*{r}Y_{\ell}^m\pqty{\hat{\mathbf{r}}},
\end{equation}

\noindent with $0\le n \le n_{\max}$, $0\le \ell \le n$ and $-\ell \le m \le \ell$. The parameter $n_{\max}$ controls the number of basis functions, $n_B = \pqty*{n_{\max} + 1}\pqty*{n_{\max} + 2}/2$, and can be adjusted according to the desired granularity of the encoding of the local environment around each atom. $Y_{\ell}^m\pqty{\hat{\mathbf{r}}}$ is a spherical harmonic, while the radial parts, $g_{n\ell}\pqty*{r}$, are built starting from the functions

\begin{equation}
    f_{n\ell}\pqty*{r} = \sqrt{\frac{1}{r_c^3}\frac{2}{u^2_{\ell n} + u^2_{\ell n+1}}}\left[\frac{u_{\ell,n+1}}{j_{\ell+1}\pqty*{u_{\ell n}}}j_{\ell}\pqty*{\frac{u_{\ell n}}{r_c}r}-
    \frac{u_{\ell n}}{j_{\ell+1}\pqty*{u_{\ell,n+1}}}j_{\ell}\pqty*{\frac{u_{\ell, n+1}}{r_c}r}\right]
    \label{eqn:radial}
\end{equation}

\noindent and executing a Gram-Schmidt orthogonalization procedure for each value of $\ell$. In Eq.~\eqref{eqn:radial}, $j_{\ell}$ stands for the $\ell$-th spherical Bessel function of the first kind and $u_{\ell,n}$ is the $\pqty*{n+1}$-th positive value at which $j_{\ell}\pqty{u} = 0$. Finally, rotational symmetry is enforced by contracting the angular parts of the projections $c_{iJn\ell m}$ of $\rho_{iJ}$ on all basis set elements $B_{n\ell m}\pqty{\mathbf{r}}$.

\begin{equation}
    p_{iJJ'n\ell} = \sum\limits_{m=-\ell}^\ell  c_{iJn\ell m}c^*_{iJ'n\ell m} =
    \frac{2\ell + 1}{4\pi}\sum\limits_{\substack{j,j'\in J,J'\\j,j' \ne i}} g_{n-\ell,\ell}\pqty*{R_{ij}}g_{n-\ell,\ell}\pqty*{R_{ij'}}P_{\ell}\pqty*{\cos\gamma_{ijj'}},
    \label{eqn:descriptors}
\end{equation}

\noindent where $\gamma_{ijk}$ is the angle defined by atoms $i$, $j$ and $j'$ (with $i$ at the vertex) and $P_{\ell}$ is the $\ell$-th Legendre polynomial. We use $p_{iJJ'n\ell}$ as our descriptors.  Therefore, no complex arithmetic is required at any point of the calculation despite the fact that the basis functions are, in general, complex.

\begin{figure}[htbp]
    \begin{center}
        \includegraphics[width=.8\columnwidth]{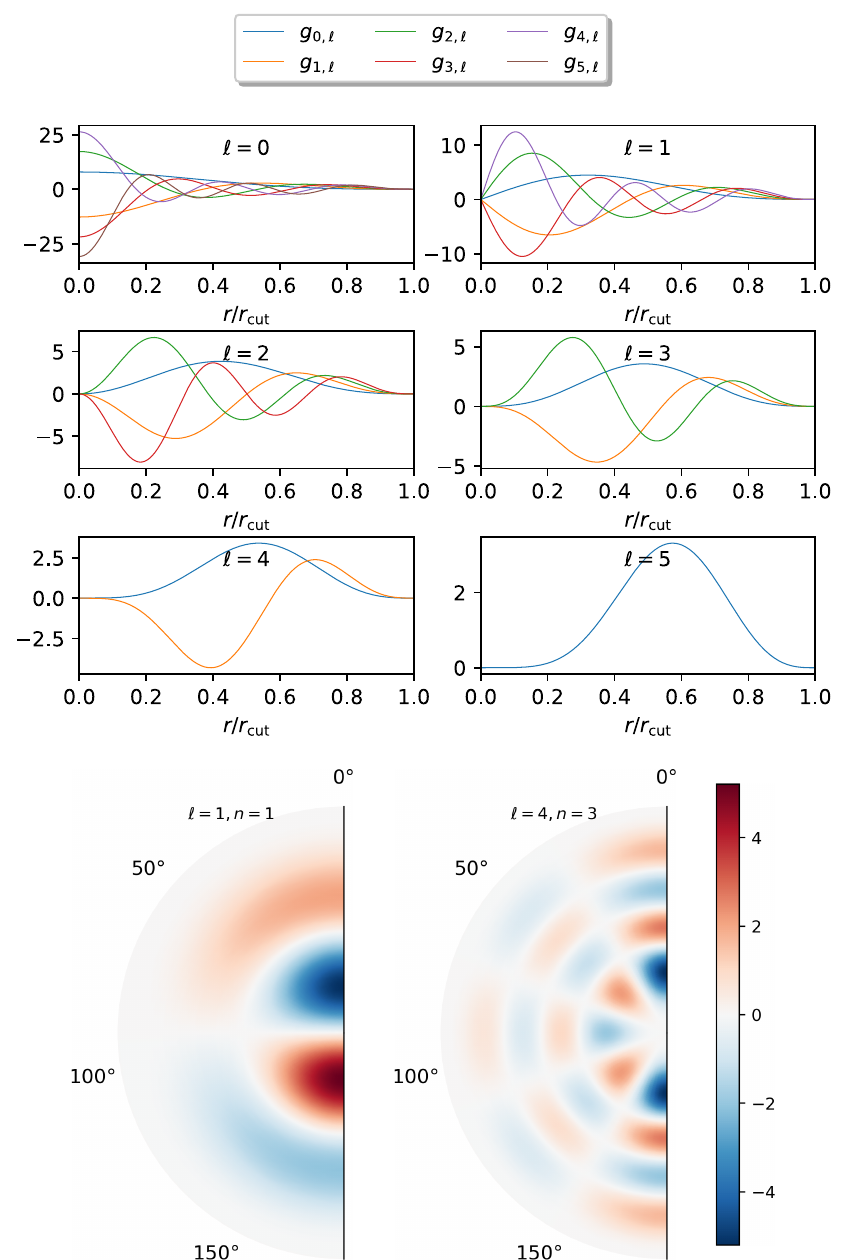}
    \end{center}
    \caption{Examples of basis functions for the spherical Bessel descriptors. Top: Radial components of all basis functions for $n_{\mathrm{max}}=5$. Bottom: Angular components of two basis functions represented in the half-plane $\phi=0$.}
    \label{fig:basis_functions}
\end{figure}

The orthogonality of the spherical harmonics and the explicit orthogonalization of the radial parts means that any pair of basis functions are orthonormal. This minimizes the amount of redundant information in the expansion and makes this choice of descriptors very compact and systematic. Furthermore, not only do the $g_{n,\ell}\pqty*{r}$ go to zero at $r=r_c$, but so do their first and second derivatives. All things considered, this scheme creates a symmetry-compatible density estimate or very smooth binning of the atomic positions around each atom. This can be readily appreciated in Fig.~\ref{fig:basis_functions}, which depicts some example basis functions: both the radial and angular parts can be regarded as creating a grid in their respective domains, with their indices determining the number of divisions of that grid.

The generalized power spectrum in Eq.~\eqref{eqn:descriptors} has
\begin{equation}
    n_p=n_Bn_{\mathrm{el}}(n_{\mathrm{el}}+1)/2
    \label{eqn:n_p}
\end{equation}
components for each atom, where $n_{\mathrm{el}}$ represents the number of distinct elements to be considered. We use $n_\mathrm{max}=4$ and with four distinct elements (C, H, O and N) in EAN, the Cartesian coordinates of the $225$ atoms in each configuration are converted into $n_{\mathrm{atoms}}\times n_\mathrm{p} = 33750$ descriptors. The choice of $n_\mathrm{max}$ is not directly connected to the number of chemical elements in the problem even though they are coincidentially the same in this instance: as mentioned above, $n_\mathrm{max}$ controls the resolution of the description of the environment in terms of both the radial and angular coordinates. Longer cutoff radii could require higher values of $n_\mathrm{max}$ to offer the same absolute spatial granularity. Since the number of descriptors increases quadratically with $n_\mathrm{max}$, a relatively low value leads to a significantly faster force field.

\begin{figure}[htbp]
    \begin{center}
        \includegraphics[width=.9\columnwidth]{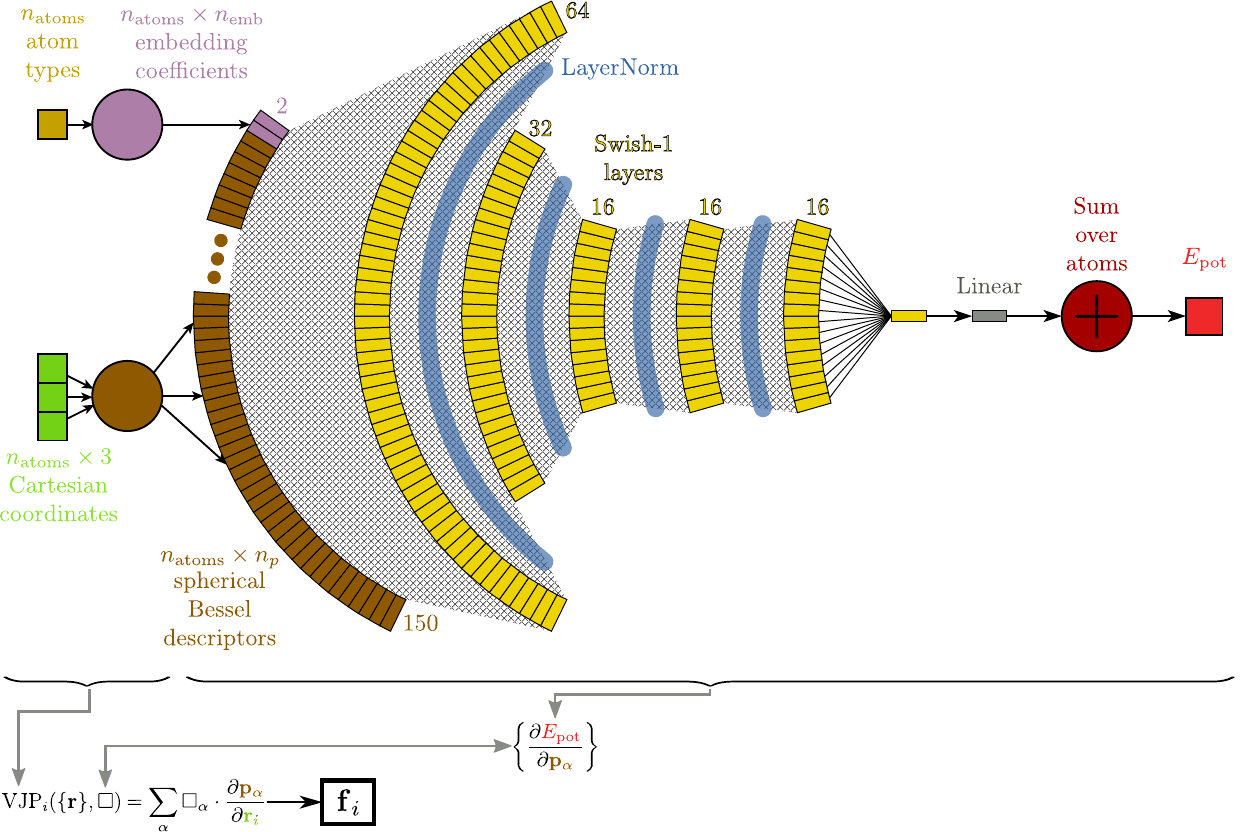}
    \end{center}
    \caption{Global schematic representation of the ML model, including the calculation of  descriptors, the embedding, and the NN. $n_{\mathrm{atoms}}$, $n_p$ and $n_{\mathrm{emb}}$ are the number of atoms, the number of descriptors [see Eq.~\eqref{eqn:n_p}] and the dimension of the embedding, respectively. The diagram at the bottom illustrates, schematically, how reverse-mode automatic differentiation computes the forces; $\alpha$ is a shorthand index that runs over all descriptors for all atoms in the system. A cross-hatch fill represents full all-to-all connectivity between adjacent layers.}
    \label{fig:full_model}
\end{figure}

The part of \textsc{NeuralIL} mapping sets of Cartesian coordinates to sets of descriptors is implemented on JAX \cite{jax2018github}, a library of composable function transformations with two key features. Firstly, it uses a just-in-time compiler to translate Python code into instructions for XLA, a highly optimized framework that improves the performance of the code by several orders of magnitude.  Through the use of that compiler, JAX aims for a different performance tradeoff than the PyTorch autograd implementation used in \textsc{TorchANI} \cite{TorchANI}, which insteads focuses on optimizing dispatch times from the Python interface. Second, JAX implements both forward- and reverse-mode automatic (also known as algorithmic) differentiation. Therefore, it is possible to obtain the explicit representations of the Jacobian or Hessian of the descriptors with respect to the Cartesian coordinates but also, more importantly, a vector-Jacobian product operator, VJP in Fig.~\ref{fig:full_model}, with a cost comparable to that of the descriptor calculation itself, a critical ingredient for the efficient calculation of the forces.

\subsection{Embedding}
\label{subsec:embedding}

The descriptors do not directly capture the chemical nature of the atom they are centered at. After training, the model can infer that piece of information indirectly because the environment of each chemical species in the IL is very characteristic and through the descriptors centered at the surrounding atoms. Still, to make our FF as general as possible with a view to its application to, for instance, alloys whose constituents are chemically similar, we also supplement the descriptors to explicitly include that piece of data. We employ the general concept of embedding, i.e., generating a low-dimensional, learned continuous representation of discrete data. This family of approaches is widely used in language processing \cite{word_embedding} and time-series analysis \cite{karingula2021boosted}. We implement it by concatenating the descriptors with the outputs of a layer taking the element index as the input and returning an array of a predefined size $n_{\mathrm{emb}}$. The elements of the vector depend only on the chemical identity of the atom and are fitted as part of the training process. Therefore, during inference the embedding layer simply supplements the descriptors with an array of predefined length from a fixed lookup table indexed by the atomic species. Increasing the dimension of the embedding, $n_{\mathrm{emb}}$, does not add any more information to the input of the neural network since the embedding array is completely determined by the chemical species at the center of each environment; however, it can impact how efficiently the model can incorporate that information. Moreover, the increase in computational cost associated to a larger $n_{\mathrm{emb}}$ is negligible. In the case of EAN, we settle on $n_\textrm{emb}=2$ because longer embedding arrays do not lead to any significant improvements in accuracy, and the input thus consists in a $n_{\mathrm{atoms}}\times \left(n_\mathrm{p}+n_{\mathrm{emb}}\right)=225\times 152$ tensor.

Embedding has been used as part of NNFFs for solid-state calculations before\cite{natalio_anharmonic_nn}, albeit in a different manner, namely by employing the embedding coefficients as weights to mix the densities of Eq.~\eqref{eqn:density} in proportions that depend on the chemical nature of the central atom. In other words, the approach of Ref.~\onlinecite{natalio_anharmonic_nn} amounts to allowing the weights in Eq.~\eqref{eq:rhoweight} to be systematically optimized and depend on the element at the center of the environment. This possibility will be discussed in more detail in the section on results.

\subsection{Neural Network Architecture}
\label{subsec:architecture}
In its most basic incarnation, a NN regression model consists in the nested application of a nonlinear activation function to a linear combination of the results of a previous activation plus a constant. This is most easily visualized in terms of a directed acyclic graph depicting the flow of data from the input to the outputs. Each yellow box in Fig.~\ref{fig:full_model} represents a neuron that receives all the outputs of all $N$ neurons from the previous layer as inputs, $\bqty*{I_i}_{i=1}^N$, and generates an output $O=f\pqty*{b + \sum\limits_{i=1}^N a_i I_i}$. Here, $f$ is the activation function, each $a_i$ is a weight and $b$ is the bias. Each neuron has its own weights and bias, and the collection of all of those make up the parameters of the model, which are chosen so as to minimize a loss function. Besides those coefficients, the flexibility of NNs lies in the choice of the activation functions, the loss, and the number and width of the layers.

A useful FF must be applicable to systems with different numbers of atoms, and to be physically sound it must also be invariant with respect to any permutation of the labels of identical atoms. We use the well known ansatz

\begin{equation}
    E_{\mathrm{pot}}\pqty*{\bqty*{\mathbf{p}_{\alpha}, \mathbf{e}_{\beta}}} =\sum\limits_{i=1}^{n_{\mathrm{atoms}}}\Omega\pqty*{\mathbf{p}_i, \mathbf{e}_i},
    \label{eqn:behler_decomposition}
\end{equation}

\noindent i.e., we consider that the energy can be decomposed into additive atomic contributions. In Eq.~\eqref{eqn:behler_decomposition}, $\Omega$ stands for the function implemented by the network (the contribution of atom $i$ to the energy), $\mathbf{p}_i$ is the collection of spherical Bessel descriptors pertaining to the environment around atom $i$, and $\mathbf{e}_i$ is the array of embedding coefficients for the same atom. Therefore, $\bqty*{\mathbf{p}_i, \mathbf{e}_i}$ is the full set of information about atom $i$ and its environment, as described in the previous sections. In contrast, $\alpha$ is a shorthand index that subsumes all indices $\pqty{i, J, J', n, \ell}$ from Eq.~\eqref{eqn:descriptors} and therefore runs over all descriptors for all atoms in the system. Likewise, $\beta$ runs over all embedding coefficients for all atoms.

Although it was introduced heuristically, this formulation has met with great success \cite{behler_tutorial}. Besides the predictive skill shown by NNs built following this template, they are easy to integrate into high-performance MD packages, whose parallelization schemes expect global reduction operations to operate on contributions to the energy and other predefined quantities from each simulation domain.

The complete \textsc{NeuralIL} model is represented schematically in Fig.~\ref{fig:full_model}, including the calculation of the spherical Bessel descriptors and the lookup of the embedding vectors. Our implementation is based on \textsc{Flax} \cite{flax}, a high-performance ML framework built on top of JAX that enables the model to be run on CPUs, GPUs and TPUs and benefit from quick and efficient automatic differentiation. The details of our final architecture are as follows. There are five hidden non-linear layers, of widths $64:32:16:16:16$. This sort of \enquote{pyramidal} architecture, with the initial layers significantly wider than subsequent ones, is found in other NN potential energy models for both molecular systems \cite{ani1} and crystals \cite{natalio_anharmonic_nn}. We chose the $64:32:16:16:16$ scheme after comparing other options found in the literature, like a shallow NN with two narrow layers of the same width \cite{singraber} and an architecture with extremely wide layers of $1000$ and $500$ neurons.\cite{natalio_anharmonic_nn} \enquote{Local} modifications, such as expanding the sequence of widths to $128:64:32:16:16:16$, does not significantly improve the results for our particular dataset. After the final non-linear layer, but before the sum over atoms, we introduce a linear layer with a single output, to account for the characteristic range and a possible offset of the potential energy.

For the nonlinear hidden layers in our networks we choose the Swish-$1$ activation function, i.e., the $\beta=1$ member of the Swish family:

\begin{equation}
    s_\beta\pqty*{x} = \frac{x}{1 + e^{-\beta  x}}.
    \label{eqn:swish}
\end{equation}

\noindent These functions are themselves the result of an automated ML-based search.\cite{swish} Like other modern activation functions (e.g., the very popular rectified linear unit, ReLU), Swish-$1$ avoids the \enquote{vanishing gradient problem} \cite{goh_hodas_vishnu} of earlier choices like the hyperbolic tangent, whereby the gradient of the loss function with respect to the weights and biases becomes vanishingly small due to the saturation of the activation functions, greatly slowing the training. However, in contrast to many of those other functions, Swish-$1$ is smooth, making it ideal for our differentiable model.

To make the training of even moderately deep and wide architectures possible and efficient, we find it essential to apply a normalization scheme between each pair of intermediate hidden  layers. As indicated in Fig.~\ref{fig:full_model}, we choose LayerNorm \cite{LayerNorm}, which centers and scales the intermediate quantities of each individual sample using their own mean and variance. LayerNorm does not require training during the forward passes and is therefore a very convenient choice for an automatically differentiable model.

\subsection{Forces}
Since the embedding coefficients do not depend on the positions, the force on atom $i$ can be computed as

\begin{equation}
    \mathbf{f}_i = -\diffp{E_{\mathrm{pot}}\pqty*{\bqty*{\mathbf{p}_{\alpha}, \mathbf{e}_{\beta}}}}{\mathbf{r}_i}=-\sum\limits_{\alpha}\diffp{E_{\mathrm{pot}}}{\mathbf{p}_{\alpha}}\cdot\diffp{\mathbf{p}_{\alpha}}{\mathbf{r}_i},
    \label{eqn:forces}
\end{equation}

\noindent In our fully differentiable model, these forces are obtained as a byproduct of the calculation of $E_{\mathrm{pot}}$ as follows. The JAX code that generates the descriptors uses reverse-mode automatic differentiation to simultaneously create the vector-Jacobian \cite{spivak1971calculus} product operator, $\VJP_i\pqty*{\bqty*{\mathbf{r}}, \square} = \sum\limits_\alpha\square_{\alpha}\cdot \diffp{\mathbf{p}_{\alpha}}/{\mathbf{r}_i}$ in Fig.~\ref{fig:full_model}, which does not depend on the NN coefficients. In our \textsc{Flax}-based implementation, this operator is seamlessly compiled together with the NN itself, which provides ${\partial E_\mathrm{pot}}/{\partial \mathbf{p}_{\alpha}}$ in Fig.~\ref{fig:full_model}, into a function that evaluates the forces in a cost- and memory-effective manner. In particular, the very large Jacobian matrix of the descriptors with respect to the atomic coordinates is never required. The total cost of the calculation is only a small and roughly constant factor higher than that of obtaining $E_{\mathrm{pot}}$ alone.

In practical terms, this means that the ML model can be trained on energies, forces or both, and used to predict energies, forces or higher-order derivatives of the energy, based on a single set of weights and biases. Since each DFT calculation yields a single energy and $3n_{\mathrm{atoms}}$ components of the forces, we find it most convenient to use only the latter. The only drawback is that the trained model is unaware of the origin of energies chosen in DFT, which is easily remedied by fitting a single constant offset under the condition that the average DFT and ML energies over the training set coincide.

\subsection{Training}

The configurations are randomly split into a training set ($90\%$ of the total) and a validation set (the remaining $10\%$). Our loss function is defined as

\begin{equation}
    \mathcal{L} = \left\langle\frac{\SI{0.1}{\electronvolt\per\angstrom}}{3n_{\mathrm{atoms}}}\sum\limits_{i=1}^{n_{\mathrm{atoms}}} \sum\limits_{C\in\left\lbrace x,y,z\right\rbrace}\log\sqty*{\cosh\pqty*{\frac{f^{\pqty*{C}}_{i,\mathrm{predicted}} - f^{\pqty*{C}}_{i,\mathrm{reference}}}{\SI{0.1}{\electronvolt\per\angstrom}}}}\right\rangle,
    \label{eqn:loss}
\end{equation}
where $\left\langle\cdot\right\rangle$ denotes an average over configurations in the current training batch. As stated above, the loss does not take into account the value of the predicted energy. This log-cosh loss \cite{losses} can be considered a smooth approximation to the mean absolute error (MAE) in the  forces. If the  prediction error $f^{\pqty*{C}}_{i,\mathrm{predicted}} - f^{\pqty*{C}}_{i,\mathrm{reference}}$ is significantly larger in absolute value than the characteristic scale parameter of \SI{0.1}{\electronvolt\per\angstrom}, its contribution to the loss is proportional to $\left\vert f^{\pqty*{C}}_{i,\mathrm{predicted}} - f^{\pqty*{C}}_{i,\mathrm{reference}} \right\vert$. On the other hand, for smaller values of the argument, $\log\sqty*{\cosh\pqty*{\Delta}}\rightarrow \Delta^2/2$, so the log-cosh can also be regarded as a robust  version of the mean square error (MSE) with built-in gradient clipping: compared to the MSE, this loss avoids an overwhelming influence on the training from  possible outliers. We found that the result of training is relatively insensitive to changes in the scale parameter that we take as \SI{0.1}{\electronvolt\per\angstrom}. It can, for example, be safely increased to \SI{1}{\electronvolt\per\angstrom}. Taking it below the expected random errors in the force predictions (e.g., to \SI{0.01}{\electronvolt\per\angstrom}) leads to the same smoothness problems posed by the MAE, while making it too large (e.g. \SI{10}{\electronvolt\per\angstrom}) makes the gradient clipping less effective and gives outliers an excessive weight in the calculated gradients. In a different application where this scale could not be guessed based on experience, a standard cross-validation approach could be used instead.

A naive implementation of Eq.~\eqref{eqn:loss} runs into overflow issues because of the exponentially increasing behavior of the hyperbolic cosine. Therefore, each contribution to the loss is actually calculated using the equivalent but more stable expression

\begin{equation}
    \frac{\log\sqty*{\cosh\pqty*{\alpha x}}}{\alpha}  = \frac{\SoftPlus\pqty*{2\alpha x} - \log 2}{\alpha} - x,
    \label{eqn:logcosh}
\end{equation}

\noindent based on the JAX implementation of $\SoftPlus{x}=\log\pqty*{1+e^x}$.

The weights of the network are initialized at random, according to a Gaussian distribution with zero mean and a standard deviation of $1/\sqrt{\text{number of inputs}}$, and the biases are initialized to zero. We minimize the loss using the adaptive moment estimation (\textsc{Adam}) algorithm \cite{ADAM} with a batch size of $8$. During the first $45\%$ of the iterations in one epoch, we increase the learning rate linearly from \num{e-3} to \num{e-2}. We then decrease it linearly back to \num{e-3} in the next $45\%$. For the last $10\%$ of the iterations we reduce the learning rate to \num{e-5}. Like in other applications of NNs, this so-called \enquote{one cycle} schedule \cite{disciplined} significantly reduces the number of epochs required to train the model, down to $500$ from the $>3000$ needed with an optimized constant learning rate of \num{4e-4}.

\section{Results and discussion}
\label{sec:results}
\subsection{Training}
\begin{figure}[htbp]
    \begin{center}
        \includegraphics[width=\columnwidth]{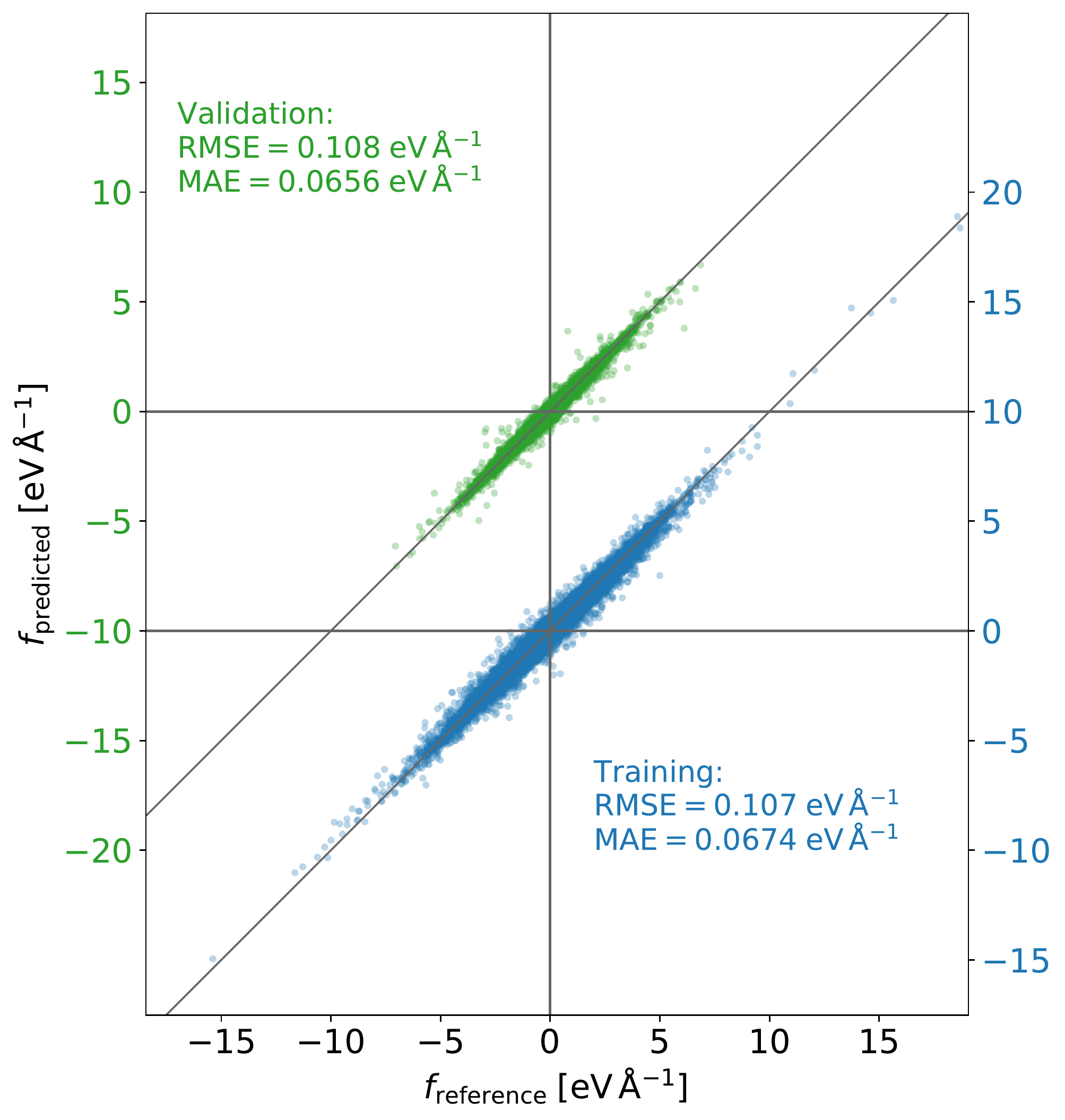}
    \end{center}
    \caption{Predicted vs. reference forces for the \textsc{NeuralIL} model over the training (blue) and validation (green) sets.}
    \label{fig:training_and_validation}
\end{figure}

Once trained to convergence, \textsc{NeuralIL} achieves a high accuracy in the prediction of forces, as evidenced by a MAE of \SI{0.0656}{\electronvolt\per\angstrom} over a validation set with a mean absolute deviation of \SI{1.11}{\electronvolt\per\angstrom}. That MAE is comparable to the differences between forces computed using different DFT implementations (e.g. LCAO vs. real-space  \cite{larsen_2008}) and can thus be described as ab-initio-like. Figure~\ref{fig:training_and_validation} shows a detailed comparison of the predicted and reference values of every component of the force on each atom in each configuration in the training and validation sets. Neither clear outliers nor particular regions with significantly worse predictions are detected. Moreover, there are no signs of overfitting during the training process or on the final result, with the validation statistics closely tracking those computed on the training set. As a first point of comparison, the accuracy of OPLS-AA as measured by its MAE with respect to DFT over the validation set for the forces is \SI{1.97}{\electronvolt\per\angstrom}, i.e, $30$ times higher. The accuracy of \textsc{NeuralIL} for predicting energies is also excellent, with a validation MAE of \SI{1.86}{\milli\electronvolt\per\atom} (or \SI{0.0429}{\kilo\calorie\per\mol}) despite the fact that energies were not included in the loss function.

The possibility of training on forces is crucial to obtaining these results with a relatively small number of configurations. For comparison, we train the same architecture on the energies of the configurations. We use the same log-cosh loss of Eq.~\eqref{eqn:loss}, but with a characteristic scale parameter of \SI{e-2}{\electronvolt\per\atom}, chosen based on arguments analogous to those presented for the forces. The learning rate schedule in this case goes from \num{e-5} to \num{e-4} and back for the first $90\%$ of each epoch before dropping to \num{e-6} for the last $10\%$. When used to predict forces, the model so created, which we call \textsc{EnergyOnly} in Tbl.~\ref{tbl:mae}, affords a MAE of \SI{0.559}{\electronvolt\per\angstrom}, rendering it unsuitable for any kind of predictive calculation. This shows that capturing the values of a function does not necessarily equate to correctly reproducing its derivatives. Interestingly, with a validation MAE of \SI{1.63}{\milli\electronvolt\per\atom}, \textsc{EnergyOnly} does not perform significantly better than \textsc{NeuralIL} when it comes to predicting energies, in keeping with the general observation that NNs trained using derivatives can achieve accuracy unmatched by those that do not take them into account \cite{NNs_with_derivatives}. Taking the total potential energy as the only piece of information describing an atomic configuration leads to a small training data set; moreover, that single piece commingles the influence of many atomic environments, leading to poor discriminatory power: indeed, the energies of all configurations visited by a system along a molecular dynamics trajectory will be distributed in a relatively narrow band compatible with the predictions of the canonical ensemble. To tackle this problem atomic decompositions of the DFT energy\cite{energy_decompositions} and local Taylor expansions of the NN energies\cite{taylor_expansion} have been devised. Training on forces, made possible by efficient automatic differentiation, makes those approximations unnecessary while achieving better accuracy. Nevertheless, it should be noted that even \textsc{EnergyOnly} outperforms OPLS-AA drastically, by a factor of $3.5$, in terms of accuracy for the forces.

In addition to its much poorer accuracy, the \textsc{EnergyOnly} model is extremely prone to overfitting. We were unable to train it below a validation energy MAE of $\sim\SI{50}{\milli\electronvolt\per\atom}$ using any fixed learning rate before that MAE started quickly diverging. Only the \enquote{one cycle} learning schedule fixed this problem.

To assess the effect of the other design features of our model, we train several alternatives, all of which are listed in Table~\ref{tbl:mae}. The \textsc{NoEmbedding} model differs from \textsc{NeuralIL} only in the fact that it lacks the embedding coefficients in the input layer. This means that the NN is in principle agnostic to the chemical nature of the atoms at the origins of each local density expressed by Eq.~\eqref{eqn:density}. However, as discussed previously, the model can still infer the element those atoms belong to from the distances and elements of the remaining atoms within the sphere. As a result, the performance \textsc{NoEmbedding} is comparable to that of \textsc{NeuralIL} for both forces and only slightly worse for energies. Since the computational cost of the embedding is negligible and  it is likely to make a bigger difference in more complicated settings, it is sensible to include it.

\begin{table}[htbp]
    \begin{tabular}{lSS}\toprule
        Model                        & {MAE ${E_{\mathrm{pot}}}$}            & {MAE $f$}                                 \\
                                     & {(\si{\milli\electronvolt\per\atom})} & {(\si{\milli\electronvolt\per\angstrom})} \\
        \midrule
        \textsc{NeuralIL}            & 1.86                                  & 65.6                                      \\
        \textsc{NoEmbedding}         & 2.26                                  & 65.7                                      \\
        \textsc{$\chi$Weights}       & 11.8                                  & 167                                       \\
        \textsc{ZWeights}            & 16.9                                  & 171                                       \\
        \textsc{WeightEmbedding}     & 7.42                                  & 109                                       \\
        \midrule
        \textsc{EnergyOnly}          & 1.63                                  & 559                                       \\
        \midrule
        \textsc{SELUActivation}      & 3.10                                  & 71.3                                      \\
        \midrule
        \textsc{AtomicCharges}       & 12.0                                  & 91.0                                      \\
        \textsc{ChargeEquilibration} & 1.93                                  & 60.8                                      \\
        \midrule
        \textsc{OPLS-AA}             & 856                                   & 1970                                      \\
        \midrule
        \textsc{DeepSets}            & 2.87                                  & 67.2                                      \\
        \bottomrule
    \end{tabular}
    \caption{Mean absolute errors (MAE) in the forces and the energies achieved with several kinds of models over the validation set. See Tbl.~\ref{tbl:description} for a short description of each model, or the main text for more extended discussion.}
    \label{tbl:mae}
\end{table}

\begin{table}[htbp]
    \begin{description}
        \item[\textsc{NeuralIL}:] The main model proposed in this article, a short-range NN potential trained on forces that combines local spherical Bessel descriptors with an embedding array determined by the chemical species and uses Swish-$1$ as its activation function.
        \item[\textsc{NoEmbedding}:] Like \textsc{NeuralIL}, but without the embedding.
        \item[\textsc{$\chi$Weights}:] Like \textsc{NoEmbedding}, but the atomic descriptors associated to the same central atom and the same two chemical species are linearly mixed using electronegativities as weights.
        \item[\textsc{ZWeights}]: Like \textsc{$\chi$Weights}, but the weights are atomic numbers instead.
        \item[\textsc{WeightEmbedding}:] Like \textsc{$\chi$Weights}, but the weights are free parameters to be optimized during training along with the other coefficients of the model.
        \item[\textsc{EnergyOnly}:] Like \textsc{NeuralIL}, but trained on total energies instead of on forces.
        \item[\textsc{SELUActivation}:] Like \textsc{NeuralIL}, but using SELU activation functions.
        \item[\textsc{AtomicCharges}:] A combination of \textsc{NeuralIL} with a Coulomb contribution to the energy and forces computed on the basis of fixed atomic charges extracted from \textsc{OPLS-AA}.
        \item[\textsc{ChargeEquilibration}:] Like \textsc{AtomicCharges}, but the atomic charges are flexible and determined using the CENT method: a second NN with the same inputs (descriptors and embedding array) as \textsc{NeuralIL} computes environment-dependent electronegativities, and the charges are calculated by solving a global optimization problem under the constraint that the system remains globally neutral.
        \item[\textsc{OPLS-AA}:] A traditional molecular-mechanics force field that has been applied to ILs, used as a baseline.
        \item[\textsc{DeepSets}:] Similar to \textsc{NeuralIL}, but atomic energies are not additive. Instead, the inputs for each atom are processed into a $16$-element array of intermediate variables, which are summed over atoms and fed to a second NN that computes the total energy.
    \end{description}
    \caption{Summary of the differences with respect to \textsc{NeuralIL} of all models discussed in this article and listed in Tbl.~\ref{tbl:mae}, for quick reference.}
    \label{tbl:description}
\end{table}

The next two models in Table~\ref{tbl:mae} are based on element-weighted descriptors $p_{in\ell}=\sum\limits_{J\le J'}\omega_{JJ'} p_{iJJ'n\ell}$ for some symmetric $\omega_{JJ'}$ and do not include any embedding vector. \textsc{$\chi$Weights} uses $\omega_{JJ'}=\chi_J \chi_{J'}$, where $\chi_J$ is the electronegativity of element $J$. The second of these models, \textsc{ZWeights}, takes $\omega_{JJ'}=Z_J Z_{J'}$, to study the effect of weighting density using atomic numbers as weights. With this change we intend to analyze how a particular choice of weights affects the discriminatory power of the model. The comparison between models that allow the NN to mix the element-specific descriptors [Eq.~\eqref{eqn:descriptors}] freely and the versions with fixed, pre-decided weights reveals how critical it is that the NN can determine and efficiently encode the type of atoms around each atomic site. Decoupling the descriptors instead of pre-mixing them improves the validation accuracy of the forces almost by a factor of three, and has an  even more marked effect on the predicted energies. Another noteworthy point is that the choice of weights is not neutral, since \textsc{$\chi$Weights} yields more accurate energies than \textsc{ZWeights}. When all possible sets of weights are considered, some are much better than others and there is a vanishing likelihood of finding a particularly good set by random chance or physical intuition, so the NN has to adapt the remaining coefficients to make up for a suboptimal choice instead. The best choice remains to keep the descriptors for different pairs of elements as separate inputs. The good accuracy of the ANI-1 FF \cite{ani1}, using a similar approach, provides additional support for this point.

Finally, \textsc{WeightEmbedding} imitates the strategy introduced in Ref.~\onlinecite{natalio_anharmonic_nn}, where the $n_{\mathrm{el}} \pqty*{n_{\mathrm{\mathrm{el}}} + 1}/2$ coefficients $\omega_{JJ'}$ form an embedding vector that is fitted during the training process and depends on the chemical species at the center of the sphere. Remarkably, the \textsc{WeightEmbedding} approach, which a priori could be expected to show good performance by introducing the information about the central atom more directly, yields a validation MAE roughly twice as high as that of \textsc{NeuralIL}. A direct cause of this drop in predictive ability may be that, even with adjustable weights, pre-mixing the densities prevents the NN from taking direct linear combinations of descriptors belonging to different element pairs and different values of $\pqty*{n,\ell}$. Another possible factor is the multiplicative effect of those weights on $p_{in\ell}=\sum\limits_{J\le J'}\omega_{JJ'} p_{iJJ'n\ell}$, the inputs to the first layer of the NN. Each change in the $\omega_{JJ'}$ affects the normalization of the inputs to all successive layers, which can be an obstacle to training. The key insight from the \textsc{WeightEmbedding} is that even an optimal choice of weights for pre-mixing descriptors corresponding to different pairs of elements, and even letting those weights depend on the central element, is not as effective a strategy as not pre-mixing the descriptors in the first place.

\subsection{The neural-network force field}

\begin{figure}[htbp]
    \begin{center}
        \includegraphics[width=.75\columnwidth]{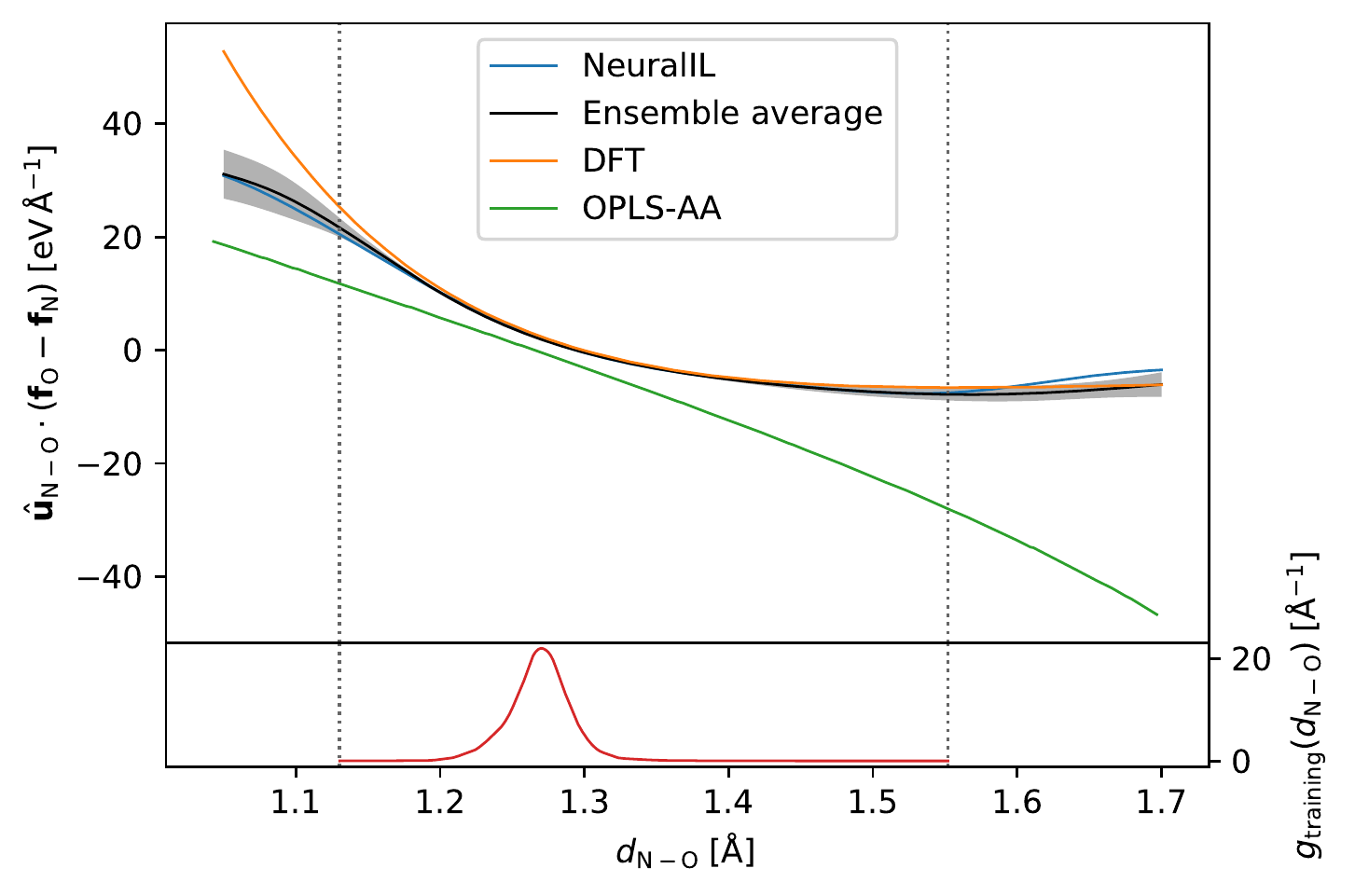}

        \includegraphics[width=.75\columnwidth]{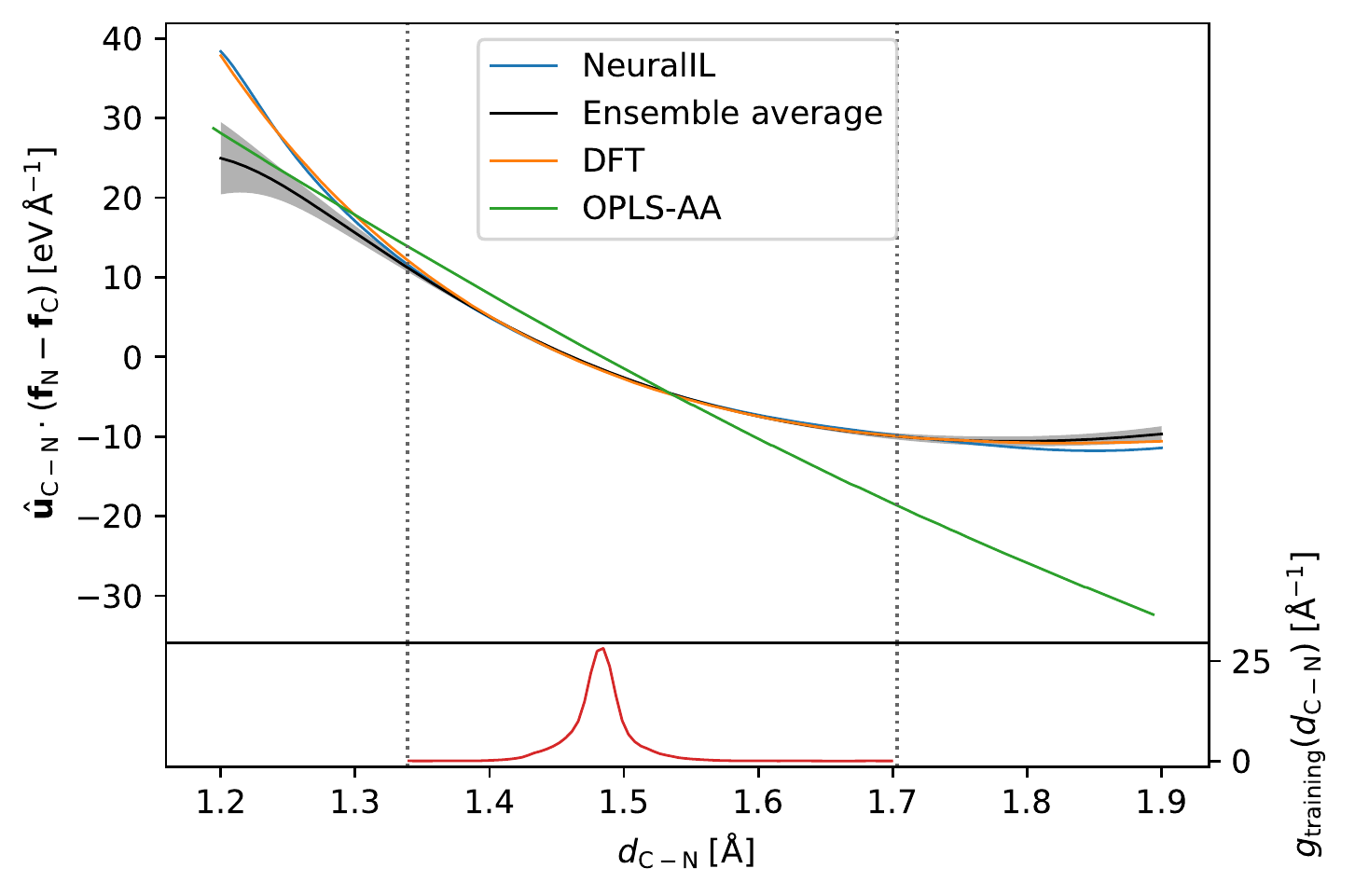}
    \end{center}
    \caption{Ensemble predictions of the projections of the N-O force (in the anion, top panel) and C-N force (in the cation, bottom panel) on the segment joining both atoms, extracted from $18$ instances of \textsc{NeuralIL} built based on random samples containing $50\%$ of the training data each. The grey area spans a single standard deviation above and below the ensemble average. Also depicted: the main \textsc{NeuralIL}, the OPLS-AA value of the same force, and the ground truth of all the NN models, i.e., the forces extracted from a \textsc{Gpaw} DFT calculation. The bottom part of each panel shows a frequency density plot of the training data for the corresponding distance. The vertical dotted lines mark the minimum and maximum values found in the training set.}
    \label{fig:bond_stretching}
\end{figure}

In contrast with molecular-mechanics force fields, our NNFF does not contain separate contributions from bond lengths, angles and dihedrals. Its parameters come from a global fit, so to evaluate the influence of a change in one of those degrees of freedom the potential energy must be evaluated along a particular trajectory that samples that deformation. The question then arises of whether this global fit leads to a loss of local detail. To explore if such a trade-off exists, we perform the following experiment. We select a random configuration from the validation set, a random anion and a random oxygen atom in it. We then displace the oxygen atom in the direction of the bond so as to change the \ce{N-O} distance, without displacing any other atom. We sample $151$ points in the interval from \SI{1.05}{\angstrom} to \SI{1.70}{\angstrom} and for each of those configurations we calculate the forces using OPLS-AA, \textsc{Gpaw} (which represents the ground truth of the NNFF) and \textsc{NeuralIL}. We perform a similar experiment with a randomly selected \ce{C-N} bond from the same configuration. The results are presented in the top and bottom panels of Fig.~\ref{fig:bond_stretching}, respectively. The OPLS-AA curves are dominated by the harmonic contribution from the stretching of each bond, with other minor bonded or non-bonded contributions that cause them to deviate from perfect straight segments. It should be noted that the most obvious point of disagreement between the DFT and OPLS-AA results, a net average offset between the corresponding force vs. distance curves, is actually a relatively trivial feature. It merely reflects the fact that the equilibrium bond lengths are different in each case and, if needed, could be corrected through a straightforward reparametrization of the OPLS-AA model. The most frequent bond lengths found in the training data are close to the OPLS-AA equilibrium value, but with an asymmetric smearing due to the contributions from the samples partially relaxed towards DFT minima. Although the LDA has a known tendency towards overbinding, compared to OPLS-AA here it seems to underestimate the equilibrium length of the \ce{C-N} bond but to overestimate that of the \ce{N-O} bond. In contrast, OPLS-AA does not afford any flexibility to solve its more fundamental discrepancies with the first-principles calculations, namely that it fails to reproduce either the local slope or the significant convexity of the force vs. distance curves. In both respects, it is clearly outperformed by \textsc{NeuralIL}, which approximates the ab-initio data accurately in a wide interval around the equilibrium bond lengths.

\subsection{Activation function}
\begin{figure}[htbp]
    \begin{center}
        \includegraphics[width=.75\columnwidth]{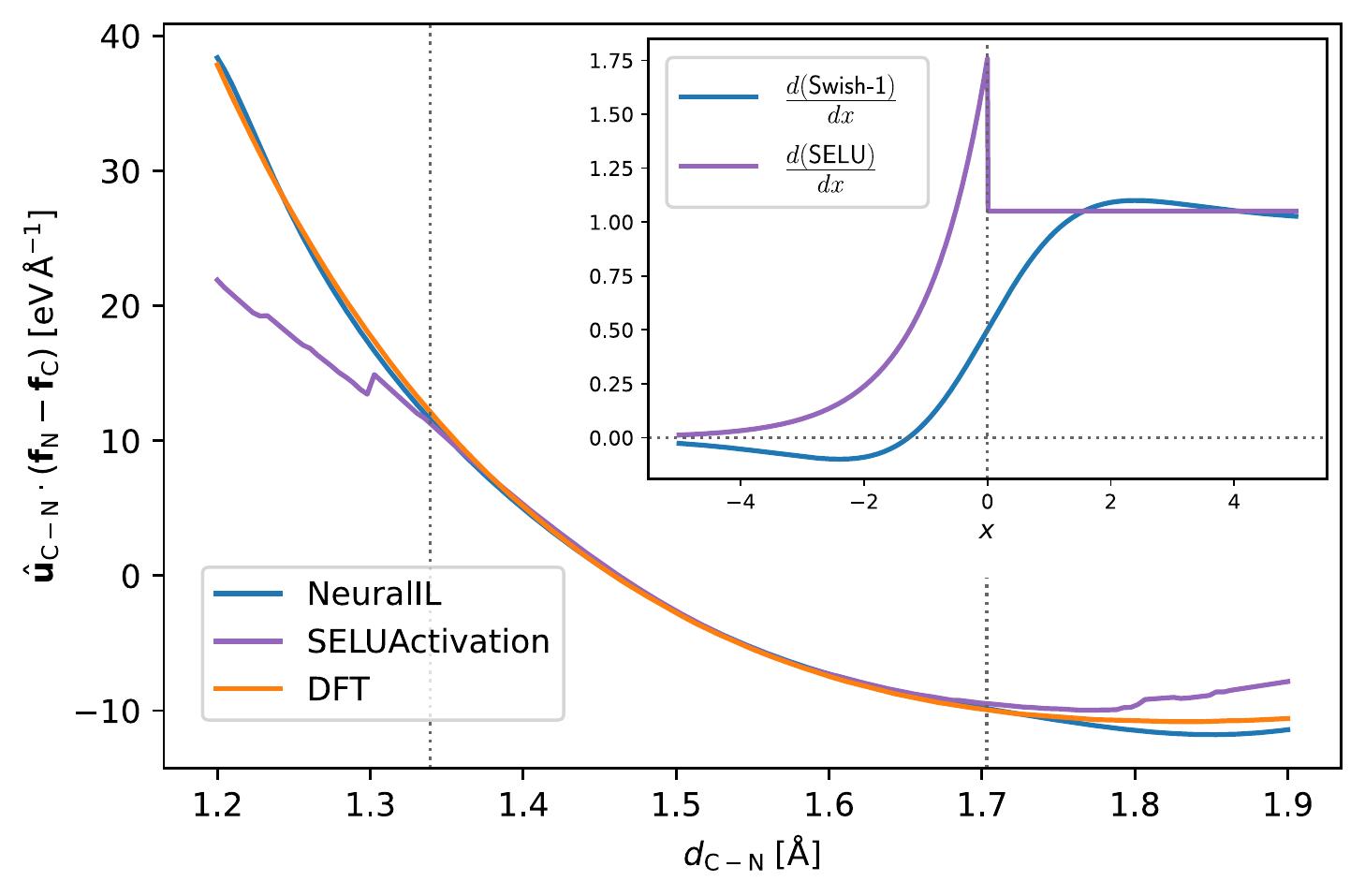}
    \end{center}
    \caption{Main panel: Comparison of the predicted projections of the C-N force \textsc{NeuralIL} on a C-N bond from \textsc{NeuralIL} and from a model identical in all respects except in that it uses the SELU instead of Swish-$1$ as the activation function. The vertical dotted lines mark the minimum and maximum values found in the training set. Inset: First derivatives of those two activation functions.}
    \label{fig:activation_functions}
\end{figure}

In our preliminary tests of different architectures we experimented with the scaled  exponential linear unit (SELU) activation function:\cite{SELU}

\begin{equation}
    \mathrm{SELU}\pqty*{x} =\begin{cases}
        \alpha\pqty*{e^x - 1}\text{, if }x\le0 \\
        \lambda x\text{, if }x>0.
    \end{cases}
    \label{eqn:SELU}
\end{equation}

\noindent The SELU was designed specifically for deep feed-forward models. We set the parameters to $\alpha=1.6732$ and $\lambda=1.0507$, carefully tuned by the original authors\cite{SELU} and shown to lead to so-called self-normalizing networks (SNNs) that naturally keep the inputs to the neurons in intermediate layers in the right range to promote fast training without additional normalization. Indeed, most of the results of this paper can be reproduced using the SELU instead of Swish-$1$ and removing the LayerNorm. However, the SELU has a discontinuity in its first derivative with some unfortunate consequences. First, the predicted forces can also have small jump discontinuities. Second, the discontinuity can be struck during training, leading to a divergence of the loss function and making the process crash. Although that problem was never observed with the $64:32:16:16:16$ architecture, the probability of triggering it increases rapidly with the number of neurons and therefore constrains the complexity of the model.

To illustrate the advantage of the Swish-$1$ activation function for an NN used to predict forces, in Fig.~\ref{fig:activation_functions} we plot the same component of the \textsc{NeuralIL}-predicted force as in the second panel of Fig.~\ref{fig:bond_stretching} together with the corresponding ground truth from DFT and with the predictions of a modified model where the activation functions have been replaced with the SELU. The SELU-based variation on \textsc{NeuralIL}, denoted as \textsc{SELUActivation} in Tbl.~\ref{tbl:mae}, performs only slightly worse than the main model. However, as Fig.~\ref{fig:activation_functions} shows, the discontinuity in its derivative introduces unphysical artifacts in the forces, especially in regions with little or no training data. The problem becomes more apparent if the depth or the width of the NN are increased, and renders this alternative architecture unsuitable for extracting higher-order derivatives of the energy, such as the Hessian. The lack of smoothnes of most modern activation functions, including the exponential linear unit (ELU) used in \textsc{TorchANI} \cite{TorchANI}, highlights the importance of physical considerations in the design of an ML regression, where the most popular choices for mainstream applications of special commercial importance like image recognition might have disqualifying features in the context of atomistic calculations.

\subsection{Ensembles}
NNFFs can, moreover, provide some indication of whether their prediction is an interpolation within an area of configuration space rich in training data, and therefore relatively safe, or an extrapolation that cannot be expected to have quantitative value. In fact, several different strategies have been proposed in the literature. Here we choose a subsampling aggregation approach\cite{ensembles}, where we train an ensemble of 18 NNs with the same architecture but each of whose training sets contain $50\%$ of the total training data, selected at random. This technique is a variation on \enquote{bagging}, which is better known for its use in the building of random-forest classification and regression models \cite{random_forests} and is made possible by the abundance of data afforded by \textsc{NeuralIL}'s use of forces. We then use each of those NNs to evaluate the forces for each of the atomic configurations described in the preceding paragraph. Figure~\ref{fig:bond_stretching} shows both the average prediction of the ensemble (as a black line) and its standard deviation (as a grey area) for each bond length. For reference, the bottom part of each panel in the figure also shows the frequency density of bond lengths in the complete training set. Looking at the standard deviations first, it is apparent that the ensemble becomes more precise in the regions where training data is abundant, which are also those where \textsc{NeuralIL} more accurately reproduces the DFT forces. The bond lengths contained in the training configurations are tightly concentrated around their most frequent values, but this region of high accuracy and precision extends well into the tails of the bond-length distributions. Interestingly, the width of the region does not seem correlated to the characteristic spread of those distributions, since the prediction for the \ce{C-N} bond remains relatively reasonable over the whole interval covered by the training data, whereas for the \ce{N-O} bond, whose lengths are less concentrated, very significant deviations are observed close to the edges of the corresponding interval. This shows the importance of the ensemble, whose spread is indeed predictive of the relative accuracy at each point. For both bonds, extrapolations beyond the boundaries of the training set are very imprecise (as measured by the standard deviation of the ensemble of NNs) and contain clear inaccuracies in most cases, with the serendipitous exception of the small-distance region for the \ce{C-N} bond. In that context, the ensemble average is also a valuable model in itself: it is not necessarily more accurate than the main model, but it is more robust with respect to outliers. In other words, it shifts the bias/variance balance towards the former in comparison with the full \textsc{NeuralIL}.

\subsection{Treatment of the electrostatic interactions}
OPLS-AA and other molecular-mechanics FFs contain electrostatic interactions in their non-bonded portions, characterized by a fixed set of atomic charges and an $r^{-1}$ dependence on the interatomic distance. Likewise, proposals to overcome the limitations of those force fields are based on more sophisticate electrostatic contributions to the energy, like those from induced dipoles. To the extent that such contributions exist, strictly short-sighted descriptors like the ones employed by \textsc{NeuralIL} cannot capture them. The short-range complexities of the interactions among atoms can be reproduced by the NN with the required flexibility regardless of their physical origin (electrostatic or otherwise), but long-range effects not correlated with the local structure cannot.

The design of \textsc{NeuralIL} deliberately omits any provision for long-range interactions to serve as a case study on how well a short-sighted FF can work for ILs. Even for simple FFs like OPLS-AA, the evaluation of the Coulomb component of the non-bonded part is a significant source of implementation complexities, especially on massively parallel environments. Since \textsc{NeuralIL} by itself delivers ab-initio-like performance, adding a long-range part to it would only be justified if that led to a significant improvement in the description of the dynamics at the atomistic level (i.e., to much more accurate energies and forces) or if it drastically reduced the error in a quantity derived from the trajectory.

Several specific methods to include electrostatic interactions in MLFFs have been proposed and demonstrated.\cite{goedecker, ceriotti, behler_4th} To carefully assess whether it is necessary, or even convenient, to include such contributions for bulk ILs, the class of system treated in this article, we analyze the results of two extremely different approaches.

Our first strategy consists in subtracting the OPLS-AA electrostatic forces (Ref.~\onlinecite{opls_parametrization}) from the DFT forces before training the NN. In other words, we combine the short-range interactions as described by the MLFF with the contribution of a system of  static atomic charges fitted to the molecular electrostatic potential at the atomic centers of the isolated gas-phase ions obtained by quantum chemical calculations at the LMP2/cc-pVTZ(-f)/HF/6-31G(d) level of theory.\cite{opls_aa_il} The resulting model, identical to \textsc{NeuralIL} in every other respect (architecture, training data, loss, learning rate schedule\ldots) is denoted as \textsc{AtomicCharges} and also included in Tbl.~\ref{tbl:mae}. Its MAEs for the energies and forces are $\sim 550\%$ and $\sim 40\%$ worse than the respective statistics for \textsc{NeuralIL}.

Those bad results do not preclude the possibility that a more sophisticated treatment could change the picture. To explore that hypothesis, we supplement \textsc{NeuralIL} with the charge equilibration via neural network technique (CENT). The method (described in Refs.~\onlinecite{goedecker,faraji_PRB_2017,behler_4th}) consists in augmenting the \textsc{NeuralIL} total energy with a term describing the electrostatic long-range interaction,
\begin{equation}
    E_{\mathrm{LR}}\pqty*{\bqty*{\mathbf{p}_{\alpha}, \mathbf{e}_{\beta}, \mathbf{r}_i}} = \sum\limits_{i=1}^{n_\mathrm{atoms}-1}\sum\limits_{j=i+1}^{n_\mathrm{atoms}} \frac{\erf{\left(\frac{r_{ij}}{\sqrt{2}\gamma_{ij}}\right)}}{r_{ij}}Q_iQ_j + \sum\limits_{i=1}^{\mathrm{n_{atoms}}}\frac{Q_i^2}{2\sigma_i\sqrt{\pi}},
    \label{eq:en_ce}
\end{equation}

\noindent where $r_{ij} = \lvert \mathbf{r}_i - \mathbf{r}_j \rvert$, the $\sigma_i$ correspond to widths of the assumed Gaussian charge density distributions of each atom (taken to be the covalent radii of the element) and $\gamma_{ij} = \sqrt{\sigma_i^2+\sigma_j^2}$. The charges, $Q_i$, are determined by a global charge equilibration scheme \cite{rappe_JPC_1991}, minimizing

\begin{equation}
    E_Q = E_{\mathrm{LR}} + \sum\limits_{i=1}^{n_{\mathrm{atoms}}}\sqty*{\chi_i\pqty*{\mathbf{p_i}, \mathbf{e}_i}Q_i + \frac{1}{2}J_i Q_i^2}
    \label{eq:ce}
\end{equation}

\noindent Here, the atomic hardnesses $J_i$ are element-specific learnable parameters, while the electronegativities $\chi_i$ are predicted by a fully connected NN similar to the one used for the short-range part, but with a $16:16:16:1$ sequence of layer widths. The overall charge conservation is enforced by introducing a Lagrange multiplier and the minimization problem is then solved with standard linear algebra routines. The short- and long-range NNs use the same descriptors and embedding coefficients as inputs, which avoids duplication of work. The CENT component is also implemented on JAX and is fully automatically differentiable. The whole model containing both neural networks is trained simultaneously to achieve the best fit to the forces. Thus, we remain close to the original CENT method and deviate from the way it is used in Ref.~\onlinecite{behler_4th}, where the NN predicting the electronegativities is trained so as to reproduce atomic charges from known configurations.

The fully trained model combining \textsc{NeuralIL} and this long-range component with fully flexible charges is denoted as \textsc{ChargeEquilibration} in Tbl.~\ref{tbl:mae}. With respect to \textsc{NeuralIL} alone, it affords a $\sim 8\%$ improvement in the validation MAE of the forces together with an insignificant degradadation in the validation MAE of the potential energy. Given that the global equilibration step couples all the atoms in the system and therefore compromises the scalability of the model, an argument can be made that the small improvement does not justify the inclusion of the CENT component for this system. This impression is reinforced by an analysis of the mean and standard deviation of the predicted charges of anions and cations: $\aqty*{q}_{\mathrm{anion}} = -0.004$, $\sigma_{{q}_{\mathrm{anion}}} = 0.016$, $\aqty*{q}_{\mathrm{cation}} = 0.001$, $\sigma_{{q}_{\mathrm{cation}}} = 0.034$ for both the training and validation sets. Typical values of those charges are therefore compatible with zero and lie orders of magnitude below the $\pm 1$ ionic charges used by OPLS-AA or the reduced $\pm 0.7$ or $\pm 0.8$ that have been used in some MD simulations of other univalent ILs.\cite{reduced_charges} The results show how the net effect of the electrostatic interaction is effectively accounted for by the short-range NN. While this at first sight might be a surprising conclusion for an IL, it agrees with a recent analysis for polarizable liquids\cite{ml_interfaces} and can be rationalized in terms of an efficient screening of electrostatic interactions in a bulk system.

As mentioned in the introduction, one of the most well known shortcomings of OPLS-AA and similar potentials is the misprediction of diffusion coefficients. Indeed, OPLS-AA describes room-temperature EAN as an almost solid ionic lattice with barely any diffusion, as evidenced by room-temperature self-diffusion coefficients of $D_{\mathrm{anion}} = \SI{1.30e-12}{\meter^2\per\second}$ and $D_{\mathrm{cation}} = \SI{6.8e-13}{\meter^2\per\second}$.\cite{ean_diffusion_md} In stark contrast, the experimentally reported diffusion coefficients are one to two orders of magnitude larger ($D_{\mathrm{anion}} = \SI{6.9e-11}{\meter^2\per\second}$ and $D_{\mathrm{cation}} = \SI{4.6e-11}{\meter^2\per\second}$).\cite{thesis_filippov} To see if \textsc{NeuralIL} overcomes these issues, we compute the diffusion coefficients. To further investigate whether it is necessary to directly include the long-range effects of polarization, we also compute those with the aforementioned \textsc{ChargeEquilibration} model. To this end, we use the JAX-MD framework\cite{jaxmd2020} to run MD simulations using each of the two models under study. We equilibrate the EAN simulation box at $T=\SI{298}{\kelvin}$ starting from the OPLS-AA trajectory and applying a Nosé-Hoover thermostat with a coupling constant $\tau_{\mathrm{NH}}=\SI{0.1}{\pico\second}$ for \SI{100}{\pico\second}. For \textsc{NeuralIL} we use an integration time step of \SI{1}{\femto\second}, while the \textsc{ChargeEquilibration} model requires a much shorter time step of \SI{0.1}{\femto\second} because of an increased tendency of the hydrogen atoms to dissociate from the rest of the cation. We then run the simulation for a further \SI{100}{\pico\second} and store the resulting trajectory to compute the temporal velocity autocorrelation function for each ion type:

\begin{equation}
    \VACF_{\substack{\mathrm{anion}\\\mathrm{cation}}}\pqty*{t}=\frac{1}{3 n_{\substack{\mathrm{anion}\\\mathrm{cation}}}}\aqty*{\sum\limits_{I=1}^{n_{\substack{\mathrm{anion}\\\mathrm{cation}}}}\mathbf{v}_I\pqty*{t_0}\cdot\mathbf{v}_I\pqty*{t_0 + t}}_T,
    \label{eqn:VACF}
\end{equation}

\noindent where $\mathbf{v}_I$ denotes the velocity of the center of mass of each ion of the corresponding type, and where the canonical ensemble average denoted by $\aqty*{\cdot}_T$ is approximated by an average over the trajectory itself. We finally use the Green-Kubo relation

\begin{equation}
    D_{\substack{\mathrm{anion}\\\mathrm{cation}}} = \int\limits_0^{\infty} \VACF_{\substack{\mathrm{anion}\\\mathrm{cation}}}\pqty*{t} \dl t
    \label{eqn:diffusion}
\end{equation}

\noindent to estimate the diffusion coefficients, and characterize its uncertainty by the oscillations of the numerical approximation to this integral in the last \SI{10}{\pico\second}. The results are $D_{\mathrm{anion}} = \SI{8.65(72)e-11}{\meter^2\per\second}$ and $D_{\mathrm{cation}} = \SI{8.24(73)e-11}{\meter^2\per\second}$ with \textsc{ChargeEquilibration}, and $D_{\mathrm{anion}} = \SI{1.00(11)e-10}{\meter^2\per\second}$ and $D_{\mathrm{cation}} = \SI{7.2(13)e-11}{\meter^2\per\second}$ with \textsc{NeuralIL}. Both models represent dramatic improvements over OPLS-AA and bring the coefficients in line with experimental measurements. The slight overestimation can be attributed to the lower density of the simulation box with respect to the actual IL at room temperature. However, the \textsc{NeuralIL} results have the advantage of capturing the $D_{\mathrm{anion}}/D_{\mathrm{cation}}$ ratio found in experiment far better, which seems to be distorted by the long-range contribution. All things considered, the fully flexible CENT term fails to add any advantageous feature to the strictly short-term \textsc{NeuralIL}. On the contrary, it hinders scalability, it requires smaller MD time steps, and it degrades the estimates of key dynamical quantities.

The conclusion is that an accurate parametrization of the potential energy of a dense ionic system does not require a specific treatment of long-range interactions. However, systems with less dense regions and correspondingly longer Debye lengths will definitely require such a treatment, and so will systems with surfaces.\cite{ml_interfaces} It is also conceivable that very specific aspects of the dynamics of a system (e.g. the frequency gap between longitudinal optical/transverse optical phonon branches in ionic solids) could hinge on particular features of the long-range interactions, but even those may be reflected in the local environment to some extent.

The non-bonded part of OPLS-AA also comprises van der Waals interactions parameterized as a $12-6$ Lennard-Jones pair potential. As outlined in our description of the classical simulations, those are truncated at a distance $\SI{6.0}{\angstrom}$. Therefore, thanks to the higher exponents of the power laws involved, those terms can still be captured by short-sighted descriptors.

\subsection{The additive ansatz}

The ansatz of additive atomic energies [Eq.~\eqref{eqn:behler_decomposition}] used by all the models discussed so far offers several important practical advantages. However, strictly speaking it can only express a subset of the permutation-invariant potential energy functions whose general form following the theory of \enquote{deep sets} \cite{deep_sets} is
\begin{equation}
    E_{\mathrm{pot}}\pqty*{\bqty*{\mathbf{p}_i, \mathbf{e}_i}} = \mu\sqty*{\sum\limits_{i=1}^{n_{\mathrm{atoms}}}\mathbf{\Omega}\pqty*{\mathbf{p}_i, \mathbf{e}_i}}.
    \label{eqn:deep_sets}
\end{equation}

\noindent There are two functions involved in the expression on the right-hand side. The first, $\mathbf{\Omega}$, maps the variables associated to each atom to a latent space where the information about the order of the inputs is destroyed (and thus permutation symmetry enforced) by a sum over atoms. The second, $\mu$, transforms the result of that sum into the potential energy. In addition to the parallelization issues, building an NN-based model for a potential energy in the form of Eq.~\eqref{eqn:deep_sets} poses several practical challenges. First, it has recently been reported \cite{deep_sets_limitations} that only a high-dimensional latent space can guarantee an adequate representation of any permutation-invariant function in practice. Second, $\mu$ must accept inputs in a broad range and still produce outputs suitable for any relevant value of $n_{\mathrm{atoms}}$, which is a bad match for the normalization techniques commonly used when training NNs. All of these problems can be avoided by restricting the output of $\mathbf{\Omega}$ to a single scalar and taking $\mu$ as the identity function or, in other words, by transferring the property of additivity from a latent space, where it is theoretically guaranteed, to the potential energy, where it becomes an approximation. Therefore, the decomposition expressed by Eq.~\eqref{eqn:behler_decomposition} is a convenient tradeoff between generality and practicality.

We now check whether switching to a more general architecture, beyond the bounds of Eq.~\eqref{eqn:behler_decomposition}, leads to a significant improvement upon the results of \textsc{NeuralIL}.
We train a new model, designated as \textsc{DeepSets} in Tbl.~\ref{tbl:mae}, and consisting of two NNs. The first one follows the scheme of \textsc{NeuralIL}, as represented in Fig.~\ref{fig:full_model}, starting from the left and up to the last $16$-neuron Swish-$1$ layer. That tensor, with $16$ components per atom, acts as the intermediate quantity denoted by $\mathbf{\Omega}$ in Eq.~\eqref{eqn:deep_sets}. Following that same equation, it is then summed over atoms and fed into the second NN that implements $\mu$ and outputs the total potential energy. That second NN is also a multilayer perceptron with Swish-$1$ as its activation function and LayerNorm between each pair of hidden layers, and with layer widths $32:32:32:1$. We use a \enquote{one cycle} training schedule that switches the learning rate from \num{e-4} to \num{e-3} and back to \num{e-4} before dropping it to \num{e-5}, and run the process for $500$ iterations just like for the rest of the models. As shown in Tbl.~\ref{tbl:mae}, the model is slightly worse than \textsc{NeuralIL} in terms of performance. On the one hand, this shows that the widely used additive ansatz expressed Eq.~\eqref{eqn:behler_decomposition} is not the only viable architecture for a fully connected feed-forward NN force field based on descriptors. On the other, it also dispels the suspicion that the ansatz is very constraining, or that dramatic boosts in accuracy are easy to obtain by generalizing it.

\section{Summary and conclusions}
\label{sec:conclusions}

We develop a neural-network-based force field for the ionic liquid ethylammonium nitrate, using forces from density functional theory as training data and modified spherical Bessel descriptors as the inputs. The validation statistics show a level of accuracy in the energies and the forces comparable to the difference between DFT implementations, and an improvement of orders of magnitude over traditional molecular-mechanics FFs like OPLS-AA, while keeping the time to evaluate the forces on a few hundreds of atoms on a single core in the milliseconds. This kind of FF can therefore be employed to calculate quantities requiring long trajectories or large samples of configurations (like thermodynamic potentials) with ab-initio accuracy. Key to its performance and flexibility is the fact that the model is automatically differentiable from end to end.

Another critical choice lies in how to include the chemical information about the system in the descriptors. We opt to describe each pair of chemical elements separately and let the neural network combine these pieces of information freely. We compare this strategy with more conventional alternatives where a set of weights, either fixed or fitted during the optimization process, is used to mix the descriptors corresponding to different elements, and show that it delivers superior results.

By training an ensemble of neural networks on random subsets of the training data, we also show how extrapolation to unexplored areas of the configuration space can be detected from the ensemble standard deviation, and how the ensemble average can provide a more robust prediction where training data is thin. This strategy can serve as a starting point to use this model as a surrogate potential energy in a first-principles calculation where generation of training data takes place on the fly and where it is important to be able to assess how reliable the prediction of the neural network is for each new configuration.

Our model represents a radical departure from the template of molecular-mechanics FFs by way of its top-bottom training process, but also in two defining features: it does not include either  a topology or a separate treatment of Coulomb interactions. However, we show that these are not obstacles to achieving both high global accuracy and a detailed description of individual degrees of freedom like bonds. This opens the door to the use of high-performance short-range potentials for this class of system where long-range electrostatic forces are traditionally considered to be critical. It should, however, be considered as valid only in the context of dense bulk systems, and not as a completely general conclusion for ionic matter.

\begin{acknowledgement}
    This work was supported by the Austrian Science Fund (FWF) (SFB F81 TACO). H. M.-C. thanks the Spanish Ministry of Education for his FPU grant.
\end{acknowledgement}

\begin{suppinfo}
    The following files are available free of charge.
    \begin{itemize}
        \item \texttt{training.json}: database of training configurations, including DFT energies and forces.
        \item \texttt{validation.json}: database of validation configurations, including DFT energies and forces.
    \end{itemize}

\end{suppinfo}

\section{Data and software availability}
\label{sec:software}
The software required to reproduce our calculations is publicly available at \url{https://bitbucket.org/sousaw/neuralil_manuscript_software} under an open-source license. The training and validation data is also included, along with the model parameters.

\providecommand{\latin}[1]{#1}
\makeatletter
\providecommand{\doi}
  {\begingroup\let\do\@makeother\dospecials
  \catcode`\{=1 \catcode`\}=2 \doi@aux}
\providecommand{\doi@aux}[1]{\endgroup\texttt{#1}}
\makeatother
\providecommand*\mcitethebibliography{\thebibliography}
\csname @ifundefined\endcsname{endmcitethebibliography}
  {\let\endmcitethebibliography\endthebibliography}{}

\end{document}